\documentclass[twocolumn,superscriptaddress,showpacs,preprintnumbers,amsmath,amssymb]{revtex4}
\usepackage{graphicx}
\usepackage{dcolumn}
\usepackage{bm}
\newif\ifpdf
\ifx\pdfoutput\undefined
\pdffalse 
\else
\pdfoutput=1 
\pdftrue
\fi
\ifpdf
\usepackage[pdftex]{graphicx}
\else
\usepackage{graphicx}
\fi
\usepackage{epsf}
\usepackage{dcolumn}
\def\be{\begin{equation}}
\def\ee{\end{equation}}
\def\bea{\begin{eqnarray}}
\def\eea{\end{eqnarray}}
\def\gsim{\ \rlap{\raise 2pt\hbox{$>$}}{\lower 2pt \hbox{$\sim$}}\ }
\def\lsim{\ \rlap{\raise 2pt\hbox{$<$}}{\lower 2pt \hbox{$\sim$}}\ }
\def\dslash{\kern-4pt \not{\hbox{\kern-2pt $\partial$}}}
\def\pslash{\not{\hbox{\kern-2pt p}}}

\def\evsq{{${\rm eV^2}$ }}
\def\L{{\rm L }}
\def\E{{\rm E }}

\def\pmutau{{${\rm P_{\mu \tau}}$ }}
\def\pmue{{${\rm P_{\mu e}}$ }}
\def\pmumu{{${\rm P_{\mu \mu}}$ }}

\def\numutonutau{{${\rm {\nu_\mu \to \nu_\tau}}$ }}
\def\numutonue{{${\rm {\nu_\mu \to \nu_e}}$ }}
\def\numutonumu{{${\rm {\nu_\mu \to \nu_\mu}}$ }}

\def\nuetonutau{{${\rm {\nu_e \to \nu_\tau}}$ }}

\begin{document}
 \ifpdf
\DeclareGraphicsExtensions{.pdf,.jpg,.mps,.png}
 \else
\DeclareGraphicsExtensions{.eps,.ps}
 \fi


\title{Large Matter Effects in \numutonutau Oscillations}


\author{Raj Gandhi}
\affiliation{
Harish-Chandra Research Institute, Chhatnag Road, Jhunsi,
Allahabad 211 019, India}

\author{Pomita Ghoshal}
\affiliation{
Harish-Chandra Research Institute, Chhatnag Road, Jhunsi,
Allahabad 211 019, India}
 
\author{Srubabati Goswami}
\affiliation{
Harish-Chandra Research Institute, Chhatnag Road, Jhunsi,
Allahabad 211 019, India}
 
\author{Poonam Mehta}
\affiliation{Department of Physics and Astrophysics, University of Delhi, 
Delhi 110 019, India}
\affiliation{
Harish-Chandra Research Institute, Chhatnag Road, Jhunsi,
Allahabad 211 019, India}

\author{S Uma Sankar}
\affiliation{
Department of Physics, I.~I.~T., Powai, 
Mumbai 400 076, India}
\date{\today}
\begin{abstract}
We show that matter effects change the 
${\rm {\nu_\mu \to \nu_\tau}}$ 
oscillation probability by as much as $70 \%$ for
certain ranges of energies and pathlengths. Consequently, the
${\rm {\nu_\mu \to \nu_\mu}}$ 
survival probability also undergoes
large changes. 	
A proper understanding of $\nu_\mu$ survival rates must 
consider matter effects in $P_{\mu \tau}$ as well as $P_{\mu e}$. 
We comment on 
a) how these matter effects may be observed and the sign of
$\Delta_{31}$ determined in atmospheric neutrino measurements
and at neutrino factories and b) how they lead to heightened
sensitivity for small $\theta_{13}$.
\end{abstract}
\maketitle

Two of the most outstanding problems in neutrino physics are 
the determination of the mixing angle $\theta_{13}$ 
\cite{footnote1} and the sign of the atmospheric 
neutrino mass-difference $\Delta_{31}$ \cite{footnote2}. 
A knowledge of  
these parameters is crucial for 
understanding the form of the neutrino mass matrix. 
So far, most studies have concentrated on the 
\numutonue oscillation probability \pmue as the means 
of determining the above parameters 
\cite{shrock}. 
This is  because the passage of neutrinos through 
earth matter dramatically changes \pmue . 

\par In this letter we point out that the 
\numutonutau oscillation probability 
\pmutau can also undergo significant change 
(a reduction as high as $\sim$ 70\% or an increase 
of $\sim$ 15\%) compared to its vacuum values over an 
observably  broad band in energies and baselines due to matter
effects. This  
can also induce appreciable changes in the  
the muon neutrino survival probability \pmumu in matter.

The muon survival rate is the primary observable
in iron calorimeter detectors like MINOS \cite{minos} 
and the proposed MONOLITH \cite{monolith} and
INO \cite{ino} and a major constituent 
of the signal in SuperKamiokande (SK) \cite{kearns},
 the planned BNL-HomeStake 
\cite{bnl-hs} large water Cerenkov detector, and several 
detectors considered for future long baseline facilities. 
The $\tau$ appearance rate as a signal for matter 
effects can also be searched for in special 
$\tau$ detectors being thought of for neutrino 
factories \cite{nufact}. 
We show that the energy ranges and baselines 
over which these effects occur are relevant 
for both atmospheric \cite{sm} and 
beam source neutrinos for the above experiments. 
Since all matter effects sensitively 
depend on the sign of $\Delta_{31}$ and on $\theta_{13}$,
 observation of the effects discussed here would provide 
information on these important unknowns.

Our discussion below uses the approximation of 
constant density and sets 
$\Delta_{21}\equiv\Delta_{\mathrm {sol}}=0$. 
Consequently the mixing angle $\theta_{12}$ and the CP phase
$\delta$ drop out of the oscillation probabilities. 
This simplifies the analytical expressions
and facilitates the qualitative discussion of matter effects. 
We have checked that this  works well (upto 
a few percent) at the energies and length 
scales relevant here.  
However, all the plots presented in this letter
are obtained by numerically solving the full three flavour 
neutrino propagation equation assuming the PREM \cite{prem} 
density profile for the earth. 
Further, these numerical calculations assume 
$\Delta_{21} = 8.2 \times 10^{-5}$ eV$^2$, 
$\sin^2\theta_{12} = 0.27$\cite{global} and $\delta=0$ \cite{footnote3}. 
We consider matter effects in neutrino probabilities only
but discuss both the cases $\Delta_{31} = \pm |\Delta_{31}|$.
We find that dramatic matter effects occur only for 
$\Delta_{31} > 0$.

\underline{{\bf 
{Review of $\mathrm{P}_{\mu {\mathrm e}}$ in matter:}}}
In vacuum, the $\nu_\mu \rightarrow \nu_e$ 
oscillation probability is 
\be
{\mathrm {P^{vac}_{\mu e}}} = 
\sin^2 \theta_{23} \sin^2 2 \theta_{13} 
\sin^2 \left(1.27 \Delta_{31} {\mathrm L}/{\mathrm E} \right),
\label{eq:pmuevac}
\ee
where 
${\mathrm{\Delta_{31}\equiv m_3^2-m_1^2}}$ 
is expressed in \evsq, \L~ in Km and \E in GeV. 
In the constant density approximation, 
matter effects can be taken into account by replacing 
$\Delta_{31}$ and $\theta_{13}$ in Eq.~(\ref{eq:pmuevac})
by their matter dependent values,
\bea
{\mathrm {
\Delta^m_{31} }} &=& 
{\mathrm {
\sqrt{(\Delta_{31} \cos 2 \theta_{13} - A)^2 +
(\Delta_{31} \sin 2 \theta_{13})^2} }}\nonumber \\
{\mathrm {\sin 2 \theta^m_{13} }}
&=& {\mathrm {\sin 2 \theta_{13} ~\Delta_{31}/
\Delta^m_{31} }}
\label{eq:dm31}
\eea
where
${\mathrm {A=2\sqrt{2}G_Fn_eE}}$
is the Wolfenstein term.
The resonance condition is 
${\mathrm {A = \Delta_{31} \cos 2 \theta_{13}}}$, 
which gives 
$E_{res} ={\mathrm{\Delta_{31} \cos 2 \theta_{13}}/2\sqrt{2}G_Fn_e}$. 
Naively, one would expect ${\mathrm {P^{mat}_{\mu e}}}$ to be maximum at 
${\mathrm{E = E_{res}}}$ since ${\mathrm{\sin 2 \theta^m_{13} = 1}}$.
But this is not true in general because at 
this energy $\Delta^m_{31}$ takes its minimum value of 
$\Delta_{31} \sin 2 \theta_{13}$ and ${\mathrm {P^{mat}_{\mu e}}}$ 
remains small for pathlengths of ${\mathrm{L \leq 1000}}$ Km. 
\begin{figure*}[htb]
\includegraphics[scale=.6]{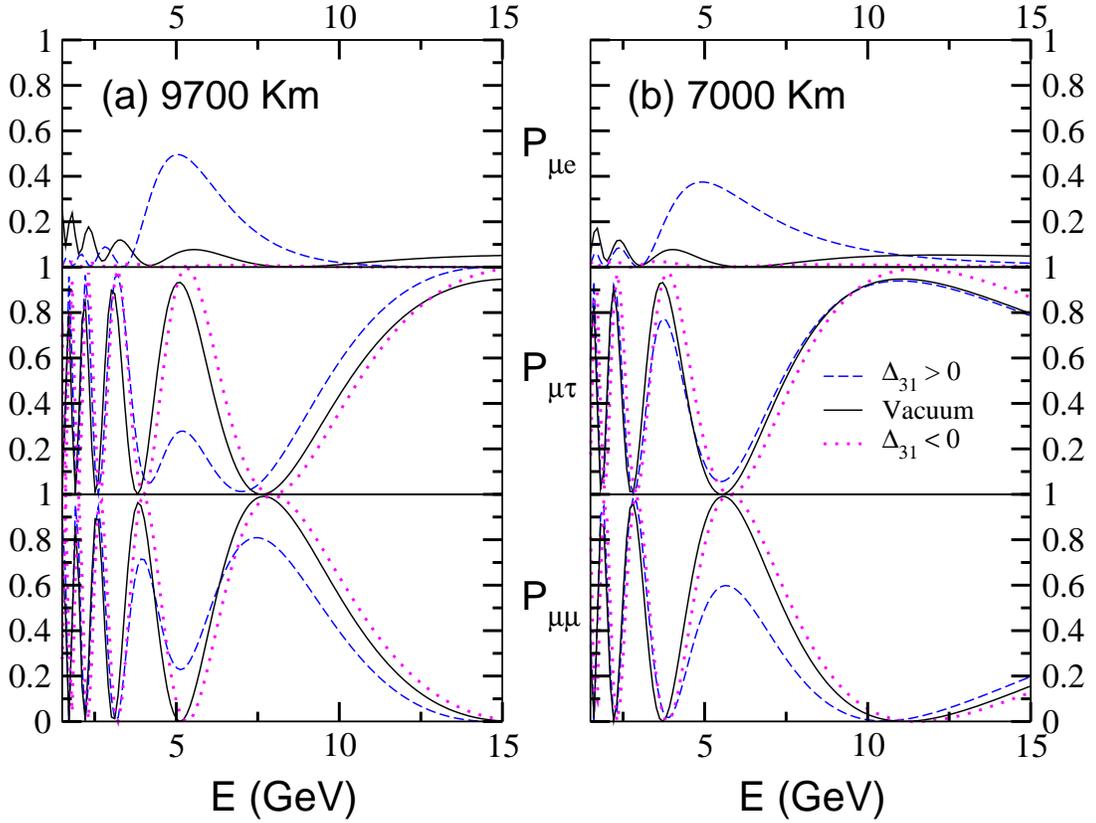}
\caption{{\em {\pmue,\pmutau and \pmumu 
plotted versus  neutrino energy,E (in GeV) 
in  matter and in vacuum for both signs of 
$\Delta m^2_{31}$ for two different baseline lengths, 
(a) for \L= 9700 Km and (b) for \L= 7000 Km. 
These plots use $\Delta_{31} = 
0.002$ eV$^2$ and $\sin^2 2\theta_{13} =0.1$.
 }}}
\label{fig:fig1}
\end{figure*}
${\mathrm {P^{mat}_{\mu e}}}$ is maximum when both 
${\mathrm {\sin 2 \theta^m_{13} = 1}}$ and 
${\mathrm {\sin^2 \left(1.27 \Delta^m_{31} L/E \right)=1
=\sin^2[(2p+1) \pi/2]}}$ are satisfied. This occurs when  
${\mathrm {E_{res} = E^{{mat}}_{{peak}}}}$. 
This gives the condition \cite{banuls}:
\be 
{\mathrm {
\rho L_{\mu e}^{max} }} \simeq
{\mathrm {
\frac{(2p+1) \pi 5.18 \times 10^3} { \tan 2\theta_{13}} ~{Km ~gm/cc} }}.
\label{eq:muecondtn}
\ee 
Here, ${\mathrm p}$ takes integer values. 
This condition is independent of $\Delta _{31}$ but depends 
sensitively on $\theta_{13}$.
Using  the product ${\mathrm {\rho_{av} L}}$ vs \L for 
the earth (calculated using the PREM profile), 
where ${\mathrm {\rho_{av}}}$ 
is the average density for a given baseline \L, 
we identify the particular values of ${\mathrm{\rho_{av} L}}$ 
which satisfy Eq.~(\ref{eq:muecondtn}) with p=0 for three different values of 
$\sin^2 2 \theta_{13}$. 
These occur at \L $\simeq$ 10200 Km, 7600 Km and 11200 Km for 
$\sin^2 2 \theta_{13} = 0.1, 0.2$ and $0.05$ respectively . 
Note that p=0 is the only relevant value of p in this case, 
given earth densities and baselines \cite{footnote4}. 
\\ 
\underline{\bf {Matter effects in \pmutau :}}
In vacuum we have
\bea
{\mathrm {P^{vac}_{\mu \tau}}} &=& 
{\mathrm {\cos^4 \theta_{13} \sin^2 2 \theta_{23} \sin^2 
\left(1.27 \Delta_{31} L/E \right)}}, 
\nonumber \\
& = & 
{\mathrm {\cos^2 \theta_{13} \sin^2 2 \theta_{23}   
\sin^2 \left(1.27 \Delta_{31} L/E \right)}} \nonumber \\
&& -~ 
{\mathrm{\cos^2 \theta_{23} P^{vac}_{\mu e}}}
\label{eq:pmutauvac} 
\eea
Including the matter effects \cite{mutaumatexp} changes this to 
\bea
{\mathrm{P^{mat}_{\mu \tau} }}& = & 
{\mathrm{
\cos^2 \theta^m_{13} {\mathrm{\sin^2 2 \theta_{23}}} 
\sin^2\left[1.27 (\Delta_{31} + A + \Delta^m_{31}) L/2E \right]}}
\nonumber \\
&+& 
{\mathrm{
\sin^2 \theta^m_{13} {\mathrm{\sin^2 2 \theta_{23}}}
\sin^2\left[1.27 (\Delta_{31} + A - \Delta^m_{31}) L/2E \right]}} 
\nonumber \\
&-& 
{\mathrm { \cos^2 \theta_{23} P^{mat}_{\mu e} }}
\label{eq:pmutaumat1}
\eea
Compared to ${\mathrm {P^{mat}_{\mu e}}}$, the 
matter dependent mass eigenstates here have a more  
complicated dependance  on the  $\nu_\mu$ and 
$\nu_\tau$  flavour content. 
Labeling the vacuum mass eigenstates as  $\nu_1$, $\nu_2$ and 
$\nu_3$, 
$\nu_1$ can be chosen to be almost entirely $\nu_e$
and $\nu_2$ to have no $\nu_e$ component \cite{fmass}.
Inclusion of the matter term ${\mathrm {A}}$ 
leaves $\nu_2$ untouched but gives a non-zero matter dependent mass 
to $\nu_1$. 
As the energy increases, the $\nu_e$ component of 
${\mathrm {\nu_1^m}}$ decreases and the  $\nu_\mu,\nu_\tau$ components 
increase such that at resonance energy they are $50 \%$. 
Similarly, increasing energy increases the ${\mathrm{\nu_e}}$
component of ${\mathrm{\nu_3^m}}$ (and reduces $\nu_\mu, \nu_\tau$ 
components) so that at resonance it becomes $50 \%$. 
Thus in the resonance region, all three matter dependent 
mass-eigenstates ${\mathrm{\nu_1^m, \nu_2^m}}$ and 
${\mathrm{\nu_3^m}}$ contain significant 
$\nu_\mu$ and $\nu_\tau$ components.
\\
We seek ranges of energy and pathlengths 
for which there are large matter effects in ${\mathrm {P_{\mu \tau}}}$,  
i.e for which  
${\mathrm{\Delta P_{\mu \tau} = P^{mat}_{\mu \tau} - P^{vac}_{\mu \tau}}}$
is large. We show that this occurs for two different sets of conditions, 
leading in one case to a decrease from a vacuum maximum and 
in another to an increase over a broad range of energies.  
\\
(i)
{\bf Large decrease in ${\mathrm{P^{mat}_{\mu \tau}}}$ 
in the resonance region :}
At energies appreciably below resonance, the 
${\mathrm{\cos^2 \theta^m_{13}}}$
term in 
Eq.~(\ref{eq:pmutaumat1}) 
$\approx {\mathrm{P^{vac}_{\mu \tau}}}$ (since
${\mathrm{\theta_{13}^m=\theta_{13}}}$, 
${\mathrm{A<< \Delta_{31}, \Delta^m_{31} 
\approx \Delta_{31}}}$) 
and the ${\mathrm{\sin^2 \theta^m_{13}}}$ term is nearly zero. 
As we increase the energy and approach resonance, 
${\mathrm{\cos^2 \theta^m_{13}}}$ begins to decrease sharply, 
while ${\mathrm{\sin^2 \theta^m_{13}}}$ increases rapidly. 
However, {\it if resonance is in the vicinity of  a vacuum peak}, 
then the decrease in the 
${\mathrm{\cos^2 \theta^m_{13}}}$ term has a much stronger impact on 
${\mathrm{P^{mat}_{\mu\tau}}}$ than the increase in the 
${\mathrm{\sin^2 \theta^m_{13}}}$ term, 
since the latter starts out at zero while the former is initially 
close to its peak value ($\approx 1$). 
As a result, ${\mathrm{P^{mat}_{\mu\tau}}}$ falls sharply. 
This fall is enhanced by the third term in Eq.~(\ref{eq:pmutaumat1}),
which is essentially 
${\mathrm{0.5 \times P^{mat}_{\mu e}}}$ 
(which is large 
due to resonance 
), 
leading to a large overall drop in ${\mathrm{P^{mat}_{\mu \tau}}}$ 
from its vacuum value. 
Note that the requirement that we be at a vacuum peak 
to begin with forces 
${\mathrm{\Delta P_{\mu \tau}}}$
to be large and negative
, with the contributions from the first and the third term 
reinforcing each other.  

The criterion for  maximal  matter effect, 
${\mathrm{E_{res} \simeq E^{{vac}}_{{peak}} }}$, 
leads to the following condition:
\be 
{\mathrm{\rho L_{\mu \tau}^{max}}} 
\simeq  
{\mathrm{(2p+1) \,\pi\, 5.18 \times 10^{3}
\,(\cos2\theta_{13}) ~{ Km~gm/cc}}}. 
\label{eq:mutaucondtn}
\ee
Unlike Eq.~(\ref{eq:muecondtn}), which has a $\tan 2 \theta_{13}$
in its denominator, 
Eq.~(\ref{eq:mutaucondtn}) has 
a much
weaker dependence on $\theta_{13}$.
This enables one to go to a higher value of ${\mathrm {p}}$
without exceeding the baselines 
relevant for observing earth matter effects. 
Incorporating the 
${\mathrm{E_{res} = E^{{vac}}_{{peak}} }}$ 
condition 
we get  
${\mathrm{\Delta P_{\mu \tau}}}$ as 
\bea
{\mathrm{\Delta P_{\mu \tau}}}  
&\simeq&  
{\mathrm {\cos^4\left[\sin2\theta_{13}(2p+1)\frac{\pi}{4}\right] - 1}}
\label{eq:delpmutau}
\eea
where we approximated $\cos 2 \theta_{13} \simeq 1$. 
We note that, in general,
${\mathrm{\Delta P_{\mu \tau}}}$
will be larger for higher
values of both ${\mathrm {p}}$ and $\theta_{13}$.
From Eq. (\ref{eq:mutaucondtn}), for p=1 and $\sin^2 2\theta_{13}
=0.1 (0.2, 0.05)$,   
${\mathrm {E_{res} = E^{vac}_{peak}}}$ occurs at 
$\sim$  9700 Km ( 9300 Km, 9900 Km ) and 
${\mathrm{\Delta P_{\mu \tau}}} \approx -0.7$ (from Eq. (\ref{eq:delpmutau})).
For p=0,  Eq. (\ref{eq:mutaucondtn}) gives
$\mathrm{ L_{\mu \tau}^{max}}$  
$\sim$ 4400 Km 
for $\sin^2 2\theta_{13} =0.1$.
However,  
${\mathrm{\Delta P_{\mu \tau}}}$ 
is 
roughly one-tenth of 
the p=1 case.
In general, for a given baseline, the choice of an optimal p is also 
dictated by the constraint that
the vacuum peak near resonance have 
a breadth which makes the effect observationally viable.

In Fig. \ref{fig:fig1}(a)  we show all three matter and vacuum
probabilities for 9700 Km.
In these plots $\Delta_{31}$ 
is taken as 0.002 eV$^2$, which gives ${\mathrm {E_{res} = E^{vac}_{peak} \sim}}$ 5 GeV. 
The middle panel of Fig. \ref{fig:fig1}(a) shows that near this energy 
${\mathrm{P^{mat}_{\mu \tau}}}$ 
($\sim$ 0.33)  
is appreciably lower compared to 
${\mathrm{P^{vac}_{\mu \tau}}}$ ($\sim$ 1).
Thus the drop due to matter effect is 
0.67, which agrees well with that 
obtained earlier using the approximate expression
Eq. (\ref{eq:delpmutau}).

\begin{figure}[htb]
\includegraphics[scale=.35]{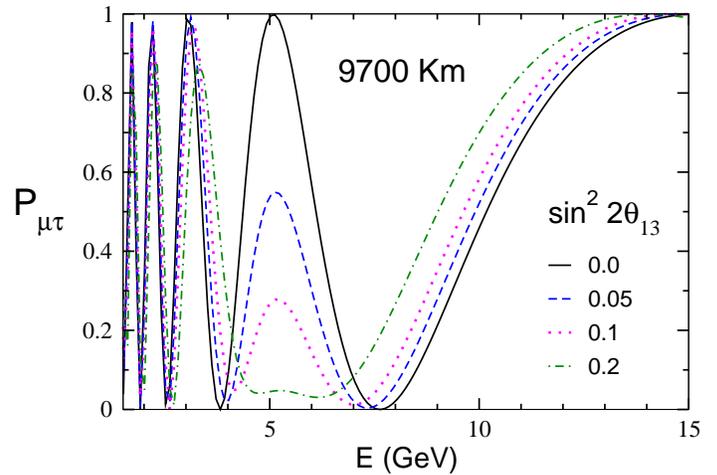}
\caption{{\em {
\pmutau plotted against the neutrino energy, E(in GeV)
for different values of $\theta_{13}$. We have used  
$\Delta_{31} =$ 0.002 eV$^2$. 
}}}
\label{fig:fig2}
\end{figure}

In Fig. \ref{fig:fig2} we show the $\theta_{13}$
sensitivity of ${\mathrm{P^{mat}_{\mu \tau}}}$ 
at 9700 Km. In particular, at ${\mathrm{E_{res} \simeq
E^{{vac}}_{{peak}}}}$ the strong dependance on $\theta_{13}$ is
governed by Eq.~(\ref{eq:delpmutau}) above.  Unlike
${\mathrm{P^{mat}_{\mu e}}}$, where the event-rate decreases as
  $\theta_{13}^2$
 for small values of $\theta_{13}$, the $\tau$ appearance rate at
${\mathrm {E_{res} = E^{vac}_{peak} }}$ increases with decreasing
$\theta_{13}$. As $\sin^2 2\theta_{13}$ goes from 0.2 to 0,
${\mathrm{P^{mat}_{\mu \tau}}}$ varies from $\sim 0.05$ to $\sim 1$. For very small values of 
$\sin^2 2 \theta_{13} (< 0.05)$ 
it will be impossible to see a {\it maximal}
resonance enhancement in $P_{\mu e}$ because the 
distance for which this occurs exceeds the diameter of the
earth. However, the observation 
of resonant suppression in $P_{\mu \tau}$ is possible, 
even for very small values of $\theta_{13}$, if the criterion,
$
N_\tau (\theta_{13} = 0) - N_\tau (\theta_{13}) \geq
3 (\sqrt{N_\tau(\theta_{13} =0)}+ \sqrt{N_\tau (\theta_{13}) }),
$
is satisfied for the tau event rate.                                                                               
 
In general the resonance has a width, and this fact 
affects observability. 
To include the width of the resonance, we write
${\mathrm {A = \Delta_{31} (\cos 2 \theta_{13} + q \sin 2 \theta_{13})}}.$ 
We find that the large matter effects discussed above still do occur 
as long as A 
is within the width of the resonance or 
$-1 \leq q \leq 1$.
\\

(ii)
{\bf Increase in \pmutau 
away from resonance :}
It is also possible for  
${\mathrm {P^{mat}_{\mu \tau}}}$ 
to differ appreciably from ${\mathrm {P^{vac}_{\mu \tau}}}$
away from resonance. 
This is evident in Fig. \ref{fig:fig1}(a) 
(central panel) in the energy range $7.5$ GeV - $ 15$ GeV. 
This effect is an enhancement 
rather than a drop, {\it i.e.}, ${\mathrm {\Delta P_{\mu \tau}}}$ 
is now positive. 
$\Delta$\pmue is small in most 
of the latter part of the energy region under 
consideration and does not contribute in an 
important way overall. 
The dominant contribution to this enhancement arises 
from the  
${\mathrm{\sin^2 \theta_{13}^m}}$ term in 
${\mathrm{{P}_{\mu\tau}}}$ 
(Eq.~(\ref{eq:pmutaumat1})) 
which is large for $E >> {\mathrm{E_{res}}}$. 
Since ${\mathrm{(\Delta_{31} + A - \Delta^m_{31}) 
\approx 2\,\Delta_{31}}}$ for these energies, 
we obtain a enhancement($\sim15\%$) which follows the vacuum curve.   
The difference between the  vacuum and matter curves 
largely reflects the difference between the $\cos^4\theta_{13}$ 
multiplicative term in the vacuum expression 
Eq.~(\ref{eq:pmutauvac}) 
and the ${\mathrm{\sin^2 \theta_{13}^m}}$ 
multiplicative term in Eq.~(\ref{eq:pmutaumat1}). While this effect is
smaller 
compared to the effect in (i) above, it occurs
over a broad energy band and may manifest itself in energy
integrated event rates.

\par Finally, we comment on the observability of the matter 
effects in ${\mathrm{P_{\mu \tau}}}$. The energies in question are above, 
but close to the $\tau$ production threshold. This suppresses 
the $\tau$  appearance rates, and will necessitate a high 
luminosity beam experiment. 
Such direct observation must perhaps await the advent of 
superbeams and/or neutrino factories.  
However, the effects  in 
${\mathrm{P_{\mu \tau}^{mat}}}$ manifest themselves indirectly in 
${\mathrm{P_{\mu \mu}^{mat}}}$, as we discuss below, and these can be observed in an 
atmospheric neutrino experiment.
\\ 
\underline{\bf{Matter effects in \pmumu :}}
The deviation of ${\mathrm{P_{\mu \mu}^{mat}}}$ 
from ${\mathrm{P_{\mu \mu}^{vac}}}$ clearly results from the 
combined effects in 
${\mathrm{P_{\mu \tau}^{mat}}}$ and ${\mathrm{P_{\mu e}^{mat}}}$ 
 {\it i.e.} 
$
{\mathrm {\Delta P_{\mu \mu} = - \Delta{P_{\mu \tau}} - 
\Delta{P_{\mu e}}}.} 
$ \\
In case (i) above, for instance, ${\mathrm{\Delta{P_{\mu \tau}}}}$ is
large and negative while
${\mathrm {\Delta P_{\mu e}}}$ is positive 
and  hence they do not 
contribute in consonance. However, 
the resulting change in \pmumu is still large, 
given the magnitude of the change ($\approx 70 \%$) 
in \pmutau. This is visible in the bottom panel of 
Fig. \ref{fig:fig1}(a), in the energy range 4-6 GeV.

One also expects a
significant drop in 
${\mathrm{P_{\mu \mu}^{mat}}}$ 
when 
either of ${\mathrm{\Delta{P_{\mu \tau}}}}$ or 
${\mathrm{\Delta{P_{\mu e}}}}$ is large and 
the other one is small.
The first of these cases (${\mathrm{\Delta{P_{\mu \tau}}}}$
large, ${\mathrm{\Delta{P_{\mu e}}}}$ small)  is shown in Fig. 
\ref{fig:fig1}(a) in the energy range 
$\sim 6-15$ GeV, with the enhancement in 
${\mathrm{P^{mat}_{\mu \tau}}}$ reflected in the decrease in 
${\mathrm{P_{\mu \mu}^{mat}}}$. 
The second case (small ${\mathrm{\Delta P_{\mu \tau}}}$, large ${\mathrm{\Delta{P_{\mu e}}}}$)  occurs when 
a minimum in the vacuum value of 
${\mathrm{P_{\mu \tau}}}$ resides in the proximity 
of a resonance, and even the rapid changes in 
${\mathrm{\sin^2 \theta_{13}^m}}$ and 
${\mathrm{\cos^2 \theta_{13}^m}}$ 
in this region fail to modify this 
small value significantly. 
This condition can be expressed as 
${\mathrm {1.27 \Delta_{31} L/E = p \pi}}$.
Note that this corresponds to a  vacuum peak of ${\mathrm {P_{\mu \mu}}}$.
Substituting E as ${\mathrm {E_{res}}}$ 
gives the distance for maximum matter effect in 
${\mathrm {P_{\mu \mu}}}$ as
\bea
{\mathrm{
\rho L_{\mu \mu}^{max} \simeq {p\,\pi\,\times10^{4} 
\,(\cos2\theta_{13})}~{Km~gm/cc}}}
\label{eq:mumucondtn}
\eea 
For p=1 this turns out to be $\sim$ 7000 Km. 
This effect \cite{mumuref} 
is 
shown in the bottom panel of Fig. \ref{fig:fig1}(b).
The large ($40\%$ at its peak) drop in \pmumu seen in this figure 
derives its strength from the resonant enhancement 
in \pmue. A sensitivity to $\theta_{13}$ around the peak similar to the one discussed
above for ${\mathrm{P^{mat}_{\mu \tau}}}$ also exists  here, leading to a larger muon survival rate as $\theta_{13}$
becomes smaller. 
\\
The width of both these effects is significant, 
ranging from $4-10$ GeV in the first case 
(Fig. \ref{fig:fig1}(b)) 
and $6-15$ GeV in the second. 
We have checked that they persist over a range of baselines 
(6000 - 9700 Km), making them observationally feasible. 
\\ 
\noindent
{\underline{\bf
{Observational Possibilities and Conclusions:} }}
We have  shown that 
large matter effects in neutrino 
oscillations are not necessarily confined to 
\numutonue  or 
\nuetonutau conversions, but can be searched for in 
\numutonutau oscillation and 
\numutonumu survival probabilities.
We have discussed their origin by studying the 
inter-relations of all the 
three matter 
probabilities, 
${\mathrm{P^{mat}_{\mu e}}}$, 
${\mathrm{P^{mat}_{\mu \tau}}}$ and 
${\mathrm{P^{mat}_{\mu \mu}}}$, and identified baseline 
and energy ranges where they act coherently to give 
observationally large effects. 
The effects discussed are strongly sensitive to the sign of 
${\mathrm{\Delta m^2_{31}}}$, 
as is apparent in the figures above. Also, there is
sensitivity 
to small $\theta_{13}$ at the energy and baseline ranges identified for
${\mathrm{P^{mat}_{\mu \tau}}}$ and 
${\mathrm{P^{mat}_{\mu \mu}}}$.
  
Specialized $\tau$ detectors operating in long baseline 
scenarios \cite{nufact}  should be able to observe effects 
like the ones discussed in the central panel of Fig. \ref{fig:fig1}(a) 
 and in Fig.  \ref{fig:fig2}. 
Similarly, detectors capable of measuring muon 
survival rates, e.g. 
magnetized iron calorimeters 
can detect the 
effects visible in the bottom panels of Fig. 
\ref{fig:fig1}(a) and (b) 
\cite{petcov}. 
 To illustrate the observability of the effect, 
we calculate that for a magnetized iron calorimeter detector 
\cite{ino, monolith} and an exposure of 
1000 kT-yr, in the energy range 
5 - 10 GeV and \L range of 6000 - 9700 Km, 
with $\Delta _{31} = +0.002 $ \evsq and 
$\sin^2 2\theta_{13} = 0.1$, 
 the total number of 
atmospheric $\mu^{-}$ events in the case of 
vacuum oscillations is 261. However,  it reduces to 204 
with matter effects.
The rates for $\mu^{+}$ in matter are identical to the 
vacuum value of 105 events.  
These numbers reflect a $4\sigma$ signal for the effect 
discussed above for ${\mathrm{P^{mat}_{\mu \mu}}}$ 
\cite{mutaudetails}.   
The Fermilab to Kamioka proposal 
\cite{9300} has a baseline of 9300 Km and is 
within the range of baselines where these effects are 
large and observable. 

Finally, we remark that although the effects
discussed here appear only for $\nu_{\mu}$ for 
$\Delta_{31}$ positive (and only for $\bar{\nu}_{\mu}$ 
for $\Delta_{31}$ negative) 
it may still be possible to
search for them in the accumulated SK data.
\\
{\bf Acknowledgments :} PM 
acknowledges 
CSIR, India for
partial financial support and  
HRI and INO for hospitality.


\begin{thebibliography}{99}
\bibitem{footnote1}
The current bound on $\theta_{13}$ is 
$\sin^2\theta_{13} < 0.05~(3\sigma)$ \cite{us}.However, this  
is  sensitive to the  value of $\Delta_{31}$ \cite{sg}.
\bibitem{footnote2}
The  best-fit value of $\Delta_{31}$ from zenith 
angle analysis of SK data is 2.1 $\times 10^{-3}$ \evsq
\cite{kearns}.
\bibitem{footnote3}We checked that for $\delta = \pm \pi/2$  
the change is within 10\%.
\bibitem{footnote4}
The approximation of ``average constant density'' 
is not a good one for core passage but is included here 
since it delineates the range of  $\sin^2 2 \theta_{13}$ where core 
passage is a requisite for maximal matter effects.
\bibitem{fmass}
While $\theta_{12}$ is 
undetermined in the approximation $\Delta_{21}=0$, it also 
drops out of the mixing matrix in this approximation, 
and the weights of $\nu_e$ in $\nu_1$ and $\nu_2$ 
do not depend on it.
\end{thebibliography}
\end{document}